\documentstyle[11pt]{article}

\begin{document}
\title{General impossible operations in quantum information}
\author{Arun K. Pati$^{(1)}$\\
Institute of Physics, Sainik School Post, \\
Bhubaneswar-751005, Orissa, India\\
$^{(1)}$ School of Informatics, University of Wales, \\
Bangor LL 57 1UT, UK}

\date{\today}
\maketitle
\def\ra{\rangle}
\def\la{\langle}
\def\ver{\arrowvert}

\begin{abstract}
We prove a general limitation in quantum information that unifies the 
impossibility principles such as no-cloning and no-anticloning. Further, 
we show that for an unknown qubit one cannot design a universal Hadamard  
gate for creating equal superposition of the original and its complement 
state. Surprisingly, we find that Hadamard transformations exist for an 
unknown qubit chosen either from the polar or equatorial great circles.
Also, we show that for an unknown qubit one cannot design a universal 
unitary gate for creating unequal superpositions of the original 
and its complement state. We discuss why it is impossible to design a 
controlled-NOT gate for two unknown qubits and discuss the implications of 
these limitations.

\end{abstract}

\vskip .5cm

PACS           NO:    03.67.-a, 03.67 Hk, 03.65.Bz\\

email:akpati@iopb.res.in\\

\vskip 1cm


\section{Introduction}

In microscopic world a qubit carries quantum as well as classical 
information. To specify the quantum
information content of an unknown qubit we need doubly infinite bits 
\cite{rj} of information, whereas to extract classical
information we need to do a measurement and that yields only a single bit 
of information. This makes a qubit so distinct from a classical bit. 
Unlike classical information there are several limitations on the 
basic operations that one can perform on quantum information. 
Using linearity of quantum evolution it can be shown that one cannot copy 
an unknown state perfectly \cite{wz,dd}.
Further, using unitarity alone it can be shown that non-orthogonal states
cannot be copied exactly \cite{hpy}.
Similarly, it was shown that there is no linear, trace preserving 
operation that takes two copies of an unknown state and 
delete a copy by acting jointly on both the copies \cite{pb,whz}. 
In addition, it was found that one cannot complement an arbitrary qubit, 
where complementing means flipping a qubit on Bloch sphere \cite{akp,bhw}. 
It was also shown that one cannot design a machine that will 
take an unknown qubit and a blank state, and produce the original along 
with a flipped state \cite{gp}. Recently, a stronger no-cloning theorem has 
been proved which says that the supplementary information needed to make a 
copy must be as much large as possible \cite{jozsa}. At the heart of these 
fundamental limitations
there lies the `unknowability' of a single quantum state.  

On the other hand there are certain type of physical operations 
that one can perform, in principle, 
on quantum information. For example, as we all know, one can swap 
an unknown state with a known or an unknown state perfectly. One can teleport 
an unknown state with the help of dual classical and quantum 
channel \cite{cb}. One can create universal entangled states of an unknown 
qubit with two types of reference states \cite{akp1} using shared 
entanglement and classical communication. One can also erase 
\cite{land,chb} the information 
content of an unknown state by swapping it with a standard state and 
then performing an irreversible operation \cite{pb}. 
Therefore, it is of utmost importance to know what are the impossible 
and possible operations on quantum information that are allowed by 
laws of quantum physics. Because these would give rise to serious 
implications for quantum computing and information processing devices in
future.

The purpose of this paper is multi fold. First, we show that there is 
no allowed transformation that will take an unknown and a blank state 
at the input port and produce 
the original along with a function of the original 
state at the output port. This limitation generalizes and unifies the 
no-cloning and no-anticloning principle for 
arbitrary qubits. Second, we show that one cannot design a Hadamard gate 
that will create a linear superposition of an unknown state along with 
its complement state with equal
amplitudes. Surprisingly, we show that there exist two distinct 
Hadamard transformations for unknown qubits chosen from the polar 
and equatorial great circles. We also show that it is not possible to design 
a unitary transformations that will create an unequal superposition of the
original qubit with its complement. Third, we show that one cannot design a 
controlled-NOT gate for two unknown qubits and discuss implications of 
these limitations. Moreover, unlike the qubits in preferred computational 
basis states, if the qubits are in some arbitrary states then the 
quantum computational logic gates cannot be designed perfectly. 

The organization of our paper is as follows. In section-II, we present our
generalized limitation. In section-III, we discuss non existence of universal
Hadamard gate and unitary gates. In section-IV, we discuss why it is
impossible to design a CNOT gate for two unknown qubits. In section-V, we
briefly discuss the implications of these limitations for future quantum 
mechanical computers and the conclusions follows.

\section{General limitation on quantum information} 

In the sequel we prove a general impossibility theorem for quantum 
information. Suppose we are given a qubit in an unknown state 
$\ver \Psi \ra= \alpha \ver 0 \ra + \beta \ver 1 \ra 
\in {\cal H}^2$, with $\alpha$, $\beta$ being {\em unknown} complex 
numbers and $\ver \alpha \ver^2 + \ver \beta \ver^2 =1$. This state is
isomorphic to any two-state system parametrized by two real parameters 
as $|\Psi(\theta, \phi) \ra= \cos \frac{\theta}{2} \ver 0 \ra + 
\sin \frac{\theta}{2} e^{i \phi} \ver 1 \ra$ with $0 \le \theta \le \pi$ and 
$ 0  \le \phi \le 2\pi$.\\

\noindent
{\bf Theorem-I:} {\em Given an arbitrary state $\ver \Psi \ra \in {\cal H}^2$ 
of an unknown qubit and a blank state $\ver \Sigma \ra \in {\cal H}{^2}$, 
there does not exist a isometric map 
${\cal M}: {\cal H}^2 \otimes {\cal H}^2\otimes {\cal H}^2
\rightarrow {\cal H}^2 \otimes {\cal H}^2 \otimes {\cal H}^2 $ that 
will transform} 
\begin{equation}
\ver \Psi \ra \otimes \ver \Sigma \ra \otimes \ver Q \ra 
\rightarrow \ver \Psi \ra 
\otimes \ver {\cal F}(\Psi) \ra \otimes \ver Q_{\Psi} \ra,
\end{equation}
where $\ver Q$ and $\ver Q_{\Psi} \ra$ are the initial and final 
states of the ancilla (it could be the corresponding states of the 
proposed machine itself).
Here $\ver {\cal F}(\Psi) \ra$ is the function of 
the original, namely, a state that is a function of $\alpha, \beta$ or their
complex conjugates. It may be related to the original state either 
by a unitary or anti-unitary transformation, i.e., 
$\ver {\cal F}(\Psi) \ra=  K \ver \Psi \ra$, 
where $K$ can be a unitary operator $U$ or anti-unitary operator $A$. 
More generally, $\ver {\cal F}(\Psi) \ra$ may be related to 
$\ver \Psi \ra$ by a sum of unitary and anti-unitary operators, i.e., 
$\ver {\cal F}(\Psi) \ra=  (\sqrt \lambda  U  + \sqrt(1 - \lambda)A)
 \ver \Psi \ra$, with $0 \le \lambda \le 1$ and $\lambda$ is real. 
Here only those unitaries and antiunitaries may be considered that gives  
isometric (only norm preserving) transformations in ${\cal H}^2$. \\

\noindent
{\em Proof:} Since a qubit in the canonical orthogonal states carry classical 
information and can be measured without any disturbance  
it can be manipulated at will. Let there be a machine that transforms a 
qubit in the orthogonal states 
$\ver 0 \ra \otimes \ver \Sigma \ra \otimes \ver Q \ra \rightarrow \ver 0 \ra 
\otimes \ver {\cal F}(0) \ra \otimes \ver Q_0 \ra$
and $\ver 1 \ra \otimes \ver \Sigma \ra \otimes \ver Q \ra 
\rightarrow \ver 1 \ra \otimes \ver {\cal F}(1) \ra \otimes \ver Q_1 \ra$. 
First, we consider the case when $K$ is
either unitary or anti-unitary.
If we send an unknown qubit through this machine, then by linearity 
we have 
\begin{eqnarray}
\ver \Psi \ra \otimes \ver \Sigma \ra \otimes \ver Q \ra &=& 
(\alpha \ver 0 \ra + \beta \ver 1 \ra) \otimes \ver \Sigma \ra \otimes 
\ver Q \ra  \nonumber\\
&\rightarrow&
\alpha \ver 0 \ra \otimes \ver {\cal F}(0) \ra \otimes \ver Q_0 \ra
+ \beta \ver 1 \ra \otimes \ver {\cal F}(1) \ra \otimes \ver Q_1 \ra.
\end{eqnarray}
and by anti-linearity of map we have
\begin{eqnarray}
\ver \Psi \ra \otimes \ver \Sigma \ra \otimes \ver Q \ra &=& 
(\alpha \ver 0 \ra + \beta \ver 1 \ra)
\otimes \ver \Sigma \ra \otimes \ver Q \ra \nonumber\\
&\rightarrow&
\alpha^* \ver 0 \ra \otimes \ver {\cal F}(0) \ra \otimes \ver Q_0 \ra
+ \beta^* \ver 1 \ra \otimes \ver {\cal F}(1) \ra \otimes \ver Q_1 \ra.
\end{eqnarray}
Note the complex conjugation on $\alpha$ and $\beta$ due to the anti-linear
nature of the map. Ideally, we should have obtained in the output port a 
state of the type 
\begin{eqnarray}
\ver \Psi \ra \otimes \ver {\cal F}(\Psi) \ra \otimes \ver Q_{\Psi} \ra &=& 
[ \alpha^2 \ver 0 \ra \otimes \ver {\cal F}(0) \ra
+ \beta^2 \ver 1 \ra \otimes \ver {\cal F}(1) \ra \nonumber\\
&+& \alpha \beta 
(\ver 0 \ra \otimes \ver {\cal F}(1) \ra
+ \ver 1 \ra \otimes \ver {\cal F}(0) \ra) ] \otimes \ver Q_{\Psi} \ra
\end{eqnarray}
when $K$ is a unitary operator or a state of the type
\begin{eqnarray}
\ver \Psi \ra \otimes \ver {\cal F}(\Psi) \ra \otimes \ver Q_{\Psi} \ra &=& 
[\ver  \alpha \ver^2 \ver 0 \ra \otimes \ver {\cal F}(0) \ra 
+ \ver \beta \ver^2 \ver 1 \ra \otimes \ver {\cal F}(1) \ra \nonumber\\
&+& \alpha \beta^* \ver 0 \ra \otimes \ver {\cal F}(1) \ra
+ \alpha^* \beta \ver 1 \ra \otimes \ver {\cal F}(0) \ra] \otimes 
\ver Q_{\Psi} \ra
\end{eqnarray}
when $K$ is an anti-unitary operator. Since the states in (2), (4)
and in (3), (5) can never be equal for arbitrary values of $\alpha$ 
and $\beta$, there is no allowed machine to satisfy (1). 
 
Next we consider the case when $|{\cal F}(\Psi) \ra$ is related to 
$| \Psi \ra$ by a sum of unitary and anti-unitary operators. 
In actuality, when we send an unknown and blank states through a machine
we will have an output state given by
\begin{eqnarray}
&& \ver \Psi \ra \otimes \ver \Sigma \ra \otimes \ver Q \ra 
\rightarrow \sqrt \lambda \alpha \ver 0 \ra \otimes U \ver 0 \ra
\otimes \ver Q_0 \ra
+ \sqrt {(1-\lambda)} \alpha^* \ver 0 \ra \otimes A \ver 0 \ra 
\otimes \ver Q_0 \ra\nonumber \\
&& + \sqrt \lambda \beta \ver 1 \ra \otimes U \ver 1 \ra 
\otimes \ver Q_1 \ra
+ \sqrt {(1-\lambda)} \beta^* \ver 1 \ra \otimes A \ver 1 \ra
\otimes \ver Q_1 \ra.
\end{eqnarray}
However, ideally we should have obtained an output state given by 
\begin{eqnarray}
&& \ver \Psi \ra \otimes \ver {\cal F}(\Psi) \ra \otimes \ver Q_{\Psi} \ra  
=[ \sqrt \lambda \alpha^2 \ver 0 \ra \otimes U \ver 0 \ra
+ \sqrt {(1-\lambda)} |\alpha|^2 \ver 0 \ra \otimes A \ver 0 \ra 
+ \sqrt \lambda \beta^2 \ver 1 \ra \otimes U \ver 1 \ra \nonumber\\
&& + \sqrt {(1-\lambda)} |\beta|^2 \ver 1 \ra \otimes A \ver 1 \ra 
 + \sqrt \lambda \alpha \beta \ver 0 \ra \otimes U \ver 1 \ra
 + \sqrt{(1- \lambda)} \alpha \beta^* \ver 0 \ra 
\otimes A \ver 1 \ra \nonumber\\
&& + \sqrt \lambda \alpha \beta \ver 1 \ra \otimes U \ver 0 \ra 
+ \sqrt {(1-\lambda)} \alpha^* \beta \ver 1 \ra \otimes A \ver 0 \ra ]
\otimes \ver Q_{\Psi} \ra.
\end{eqnarray}
Since (6) and (7) can never be the same for arbitrary values of 
$\alpha$ and $\beta$, we conclude that the generalized
machine does not exist for an unknown qubit. Hence the proof.

The non existence of a machine defined in (1)
is a class of general form of limitations that one can impose on 
quantum information. Some known impossible machines can be thought of as 
special cases of the above impossible machine. For example, if 
$\ver {\cal F}(\Psi) \ra = \ver 
\Psi \ra$, then it is the no-cloning principle, as the unitary
operator $K = I$, with $I$ being the identity operation. If $\ver {\cal F}
(\Psi) \ra = \ver \Psi^* \ra = \alpha^* \ver 0 \ra + \beta^* \ver 1 \ra=
{\cal C} \ver \Psi \ra$, with ${\cal C}$ being conjugation operation, 
then this limitation suggests that starting with an unknown qubit it
is impossible to produce the original and a conjugate qubit.
Here, $K$ will be the 
conjugating operation which is an anti-unitary operator. If $\ver 
{\cal F}(\Psi) \ra = \ver {\bar \Psi} \ra$, where
$\ver {\bar \Psi} \ra = \alpha^* \ver 1 \ra - \beta^* \ver 0 \ra$ 
then $K$ is flipping operation and is conjugating up to 
a unitary operator. In this case our limitation becomes impossibility of 
producing a complement copy along with the original starting from a single
copy. This can be regarded as a new limitation on quantum information. 
Note that it is not same as no-complementing principle which states 
that the operation $\ver \Psi \ra \rightarrow \ver {\bar \Psi} \ra$ is an
impossible operation \cite{akp,bhw}. In the present case, it aims to preserve
the original and produce a complement copy and that is an impossible one. 
Since any anti-unitary transformation is conjugating 
times unitary transformation, one can relate the complement and conjugate 
states for a qubit as $\ver {\bar \Psi} \ra = (- i \sigma_y) {\cal C} 
\ver \Psi \ra$. Thus, we are able to find new limitations as well as
unify three principles under a general impossible machine. 

When $K$ is a sum of unitary and anti-unitary transformation
then we have a new type of impossible machine and it 
becomes very interesting indeed. For example if $U=I$ and $A$ is 
complementing operation, then the transformation (1) will suggest 
\begin{equation}
\ver \Psi \ra \otimes \ver \Sigma \ra \otimes \ver Q \ra \rightarrow 
[\sqrt \lambda \ver \Psi \ra \otimes \ver \Psi \ra + 
\sqrt{(1-\lambda)} \ver \Psi \ra \otimes \ver {\bar \Psi} \ra]
\otimes \ver Q_{\Psi} \ra
\end{equation}
which can be called an impossible ``{\it cloning-cum-complementing}'' 
quantum machine. Because when $\lambda =1$ it will be purely a 
quantum cloning and 
when $\lambda =0$
it will be purely quantum complementing machine. For any intermediate value of 
$\lambda$ the machine will be a hybrid one. Since we cannot have an exact
hybrid machine, it would be very interesting to
see how the optimal values of the fidelity for such an approximate machine
behaves as a function of the known parameter $\lambda$. Here fidelity may be
defined in the usual sense as the overlap of the ideal output with the actual
output state (in general a mixed state) $\rho_{\rm actual}$, i.e., 
$F=\la {\cal F}(\Psi) \ver 
\rho_{\rm actual} \ver {\cal F}(\Psi) \ra$. However our
purpose is not to study approximate machines, but to {\em discover} new 
physical operations that cannot be done exactly. We can suggest that if in 
future one discovers some other limitations, then those may be encompassed 
by our new principle. One may notice that the quantum copy-deleting machine 
proposed in \cite{pb} does not belong to the above class of machines
because the deletion operation maps 
$\ver \Psi \ra \otimes \ver \Psi \ra \otimes \ver Q \ra 
\rightarrow \ver \Psi \ra \otimes \ver \Sigma \ra \otimes \ver Q_{\Psi} \ra$.

\section{Non existence of universal Hadamard and Unitary gates}

In this section we discuss two other limitations that does not belong to the 
above class.  We prove that it is impossible to design 
some important one-qubit gates for a qubit in some unknown state. 
First, we show why it is impossible to have a Hadamard gate in a 
universal way. Second, we show that one cannot design a unitary gate that 
will create unequal superposition of unknown state with its complement.

It is beyond doubt that in quantum computation and 
information theory two ubiquitous gates are Hadamard and CNOT. 
These gates are very useful in various
quantum algorithms (like Deutsch-Jozsa, Shor, and Grover etc.) and information 
processing protocols \cite{nc}. We will prove that one {\em cannot design 
these useful logic gates for arbitrary, 
unknown qubits}. We know that if we are given a qubit in either 
$\ver 0 \ra$ or $\ver 1 \ra$ state, then the Hadamard transformation 
(one-qubit gate) rotates
a qubit in the state $\ver 0 \ra \rightarrow {1 \over \sqrt 2}(\ver 0 \ra +  
\ver 1 \ra )$ and $\ver 1 \ra \rightarrow {1 \over \sqrt 2}
(\ver 0 \ra - \ver 1 \ra )$, i.e., it creates superposition of the
original and its complement state with equal amplitudes. The question 
is if we are given an
unknown qubit pointed in some arbitrary direction ${\bf n}$ in a state 
$\ver \Psi \ra$ or in the direction $-{\bf n}$ in a state $\ver {\bar \Psi} 
\ra$ can we design a logic gate that will transform these inputs as follows:
\begin{eqnarray}
&& \ver \Psi \ra \rightarrow {1 \over \sqrt 2}(\ver \Psi \ra +  
\ver {\bar \Psi}  \ra ) \nonumber\\ 
&& \ver {\bar \Psi}  \ra \rightarrow {1 \over \sqrt 2}(\ver \Psi  \ra - 
\ver {\bar \Psi}  \ra ), 
\end{eqnarray}
where one can imagine that one half of the Bloch sphere has been chosen to
play the role of $\ver \Psi\ra$ and the other half to play the role of
$\ver {\bar \Psi}  \ra $. Alternately, a naturally universal way of defining
a Hadamard gate would be
\begin{eqnarray}
&& \ver \Psi \ra \rightarrow {1 \over \sqrt 2}(\ver \Psi \ra +  
i \ver {\bar \Psi}  \ra ) \nonumber\\ 
&& \ver {\bar \Psi}  \ra \rightarrow {1 \over \sqrt 2}( i \ver \Psi  \ra + 
\ver {\bar \Psi}  \ra ).
\end{eqnarray}
The later definition has an advantage that the transformation is invariant
if we interchange $\ver \Psi \ra $ and $\ver {\bar \Psi}  \ra $.
But as we will see subsequently, both the definitions have their own advantages
when applied to special classes of unknown qubits.

{\bf Theorem-II:} {\em There is no Hadamard gate defined by (9) or (10) 
for an unknown qubit that will create an equal superposition of the 
original state $|\Psi \ra$ and its complement state $|{\bar \Psi} \ra$.}

We can prove this using either the linearity 
of quantum evolution or the unitarity. The proof below is based on the 
unitarity.\\

\noindent
{\bf Proof}: 
Suppose that there exist a universal Hadamard gate for all possible 
inputs chosen from Bloch sphere. If it is so, then for any two distinct 
qubits $\{\ver \Psi_1 \ra,
\ver \Psi_2 \ra \}$ and their complement states $\{\ver {\bar \Psi_1} \ra,
\ver {\bar \Psi_2} \ra \}$, by (9) we must have
\begin{eqnarray}
&& \ver \Psi_1 \ra \rightarrow {1 \over \sqrt 2}(\ver \Psi_1 \ra +  
\ver {\bar \Psi_1}  \ra ) \nonumber\\ 
&& \ver {\bar \Psi_1}  \ra \rightarrow {1 \over \sqrt 2}(\ver \Psi_1 \ra - 
\ver {\bar \Psi_1}  \ra ). 
\end{eqnarray}
And similarly, we must have
\begin{eqnarray}
&& \ver \Psi_2 \ra \rightarrow {1 \over \sqrt 2}(\ver \Psi_2  \ra +  
\ver {\bar \Psi_2}  \ra ) \nonumber\\ 
&& \ver {\bar \Psi_2}  \ra \rightarrow {1 \over \sqrt 2}(\ver \Psi_2 \ra - 
\ver {\bar \Psi_2}  \ra ). 
\end{eqnarray} 
Now taking the inner product, we have
\begin{eqnarray}
\la \Psi_1 \ver \Psi_2 \ra &\rightarrow&  \frac{1}{2}
(\la \Psi_1 \ver \Psi_2 \ra + \la \Psi_1 \ver {\bar \Psi_2} \ra
+ \la {\bar \Psi_1} \ver \Psi_2 \ra + \la {\bar \Psi_1} \ver 
{\bar \Psi_2} \ra ) \nonumber\\
\la {\bar \Psi_1} \ver {\bar \Psi_2} \ra &\rightarrow&  \frac{1}{2}
(\la \Psi_1 \ver \Psi_2 \ra - \la \Psi_1 \ver {\bar \Psi_2} \ra
- \la {\bar \Psi_1} \ver \Psi_2 \ra + \la {\bar \Psi_1} \ver 
{\bar \Psi_2} \ra ). 
\end{eqnarray}

Similarly, if we consider the Hadamard transformation defined by (10) then
for two arbitrary qubits we have the inner product condition
\begin{eqnarray}
\la \Psi_1 \ver \Psi_2 \ra &\rightarrow&  \frac{1}{2}
(\la \Psi_1 \ver \Psi_2 \ra + i \la \Psi_1 \ver {\bar \Psi_2} \ra
- i \la {\bar \Psi_1} \ver \Psi_2 \ra + \la {\bar \Psi_1} \ver 
{\bar \Psi_2} \ra ) \nonumber\\
\la {\bar \Psi_1} \ver {\bar \Psi_2} \ra &\rightarrow&  \frac{1}{2}
(\la \Psi_1 \ver \Psi_2 \ra - i\la \Psi_1 \ver {\bar \Psi_2} \ra
+ i \la {\bar \Psi_1} \ver \Psi_2 \ra + \la {\bar \Psi_1} \ver 
{\bar \Psi_2} \ra ). 
\end{eqnarray}

Taking $\ver \Psi_i \ra = \alpha_i \ver 0 \ra + \beta_i \ver 1 \ra$ and
$\ver {\bar \Psi_i} \ra = \alpha_i^* \ver 1 \ra - \beta_i^* \ver 0 \ra$
with $i=1,2$, we can check that for two arbitrary qubits 
$\la \Psi_i \ver {\bar \Psi_j} \ra
= -\la {\bar \Psi_i} \ver \Psi_j \ra^*$ and  $\la \Psi_i \ver \Psi_j \ra
= \la {\bar \Psi_i} \ver {\bar \Psi_j} \ra^*$ is always satisfied. With 
these conditions, it is clear that the inner product is not preserved. 
Hence a universal Hadamard gate defined by (9) or (10) cannot exist for 
arbitrary qubits. In quantum interferometric language {\em one cannot design
a $50/50$ beam splitter for an unknown photon} that creates a equal
superposition of photon polarization with its orthogonal counterpart.
This is a very important limitation as it suggests that linearity does
not allow us to linearly superpose an unknown state with its complement.
 
One may wonder are there any special class of qubits for which universal
Hadamard gate exist? It may be remarked that even though it is not 
possible to flip an arbitrary qubit, a qubit chosen from equatorial 
or polar great circle on a Bloch sphere can be flipped exactly 
\cite{akp2}. This is also the largest set of states
on Bloch sphere that can be complemented perfectly\cite{sb}. 
Surprisingly, and somewhat curiously, here we will show that if we 
restrict our qubits from polar great circle then there {\em exists}
Hadamard transformation (9) for {\em unknown} values of $\theta$, 
but not for qubits from equatorial great circle. 
If we restrict our qubits from equatorial great circle then there {\em exists} 
Hadamard transformation (10) for {\em unknown} $\phi$, but not for qubits 
from polar great circle.

With the computational basis of a qubit, if $|0\ra$ represents a point
on the north pole and  $|1\ra$ represents a 
point on the south pole $|1 \ra$, then the union of the sets ${\cal S}_P^+ U 
{\cal S}_P^-$ represents polar great circle, where ${\cal S}_P^+ :=
\{ \ver \Psi(\theta) \ra ~~|~~ \ver \Psi(\theta) \ra = 
\cos \frac{\theta}{2} \ver 0 \ra + 
\sin \frac{\theta}{2}\ver 1 \ra, 0 \le \theta \le \pi \}$ and 
${\cal S}_P^- :=
\{ \ver {\bar \Psi}(\theta) \ra ~~|~~ \ver {\bar \Psi}(\theta) \ra = 
\cos \frac{\theta}{2} \ver 1 \ra - 
\sin \frac{\theta}{2}\ver 0 \ra, 0 \le \theta \le \pi \}$. Similarly, the 
union of the sets
${\cal S}_E^+ U {\cal S}_E^-$ represents equatorial
great circle where ${\cal S}_E^+ := \{\ver \Psi(\phi) \ra ~~|~~ 
\ver \Psi(\phi) \ra=
\frac{1}{\sqrt2}( |0 \ra + e^{i\phi}| 1 \ra), 0  \le \phi \le 2\pi \}$ 
and 
${\cal S}_E^- := \{\ver \Psi(\phi) \ra ~~|~~ 
\ver \Psi(\phi) \ra=
\frac{1}{\sqrt2}( |1 \ra - e^{-i\phi}| 0 \ra), 0  \le \phi \le 2\pi \}$.
These class of qubits belong to one dimensional subspace of $S^2$ and 
play a very special role because these are the ones which can also be 
remotely prepared using one unit of quantum entanglement and one bit of 
classical communication \cite{akp2}. This gives a hint that may 
be for these class of qubits one can design Hadamard gates. 

First, consider the Hadamard transformation defined in (9). 
The reason why a Hadamard gate (9) exist for the polar great circle 
is that it preserves the inner product
condition (13). One can check that for this set if we denote 
$\ver \Psi_1\ra= \ver \Psi(\theta_1) \ra $ and
$\ver \Psi_2\ra= \ver \Psi(\theta_2) \ra $ and so on, then one has 
\begin{eqnarray}
&& \la \Psi(\theta_1) \ver {\bar \Psi}(\theta_2) \ra
= -\la {\bar \Psi}(\theta_1) \ver \Psi(\theta_2) \ra \nonumber\\   
&& \la \Psi(\theta_1) \ver \Psi(\theta_2) \ra
=\la {\bar \Psi}(\theta_1) \ver {\bar \Psi}(\theta_2) \ra
\end{eqnarray} 
for arbitrary non-zero values of $\theta$. 
This crucial condition ensures that the unitarity (13) is {\em not violated} 
for polar qubits. However, if we take qubits from equatorial great circle, 
then any qubit and its complement can be written as $|\Psi(\phi) \ra = 
H(\cos \frac{\phi}{2} \ver 0 \ra 
- i \sin \frac{\phi}{2} \ver 1 \ra)$ and 
$|{\bar \Psi}(\phi) \ra = H(i \sin \frac{\phi}{2} \ver 0 \ra
- \cos \frac{\phi}{2} \ver 1 \ra )$ up to an overall phase, where $H$ is
the ordinary Hadamard gate. One can check that the following conditions hold
for equatorial qubits:
\begin{eqnarray}
\la \Psi(\phi_1) \ver {\bar \Psi}(\phi_2) \ra
&=& \la {\bar \Psi}(\phi_1) \ver \Psi(\phi_2) \ra \nonumber\\
\la \Psi(\phi_1) \ver \Psi(\phi_2) \ra
&=& \la {\bar \Psi}(\phi_1) \ver {\bar \Psi}(\phi_2) \ra.
\end{eqnarray}
With this condition the inner product condition (13) is not preserved
and hence there cannot be a Hadamard gate (9) for equatorial great circles.

Second, consider the Hadamard transformation defined by (10). 
One can check that if we chose qubits from polar great circle then
using conditions (15), the unitarity condition (14) {\em is violated}.
But for qubits chosen from equatorial great circle, using condition (16),
unitarity requirement (14) is satisfied. Hence one can design a 
Hadamard gate defined by
(10) for equatorial qubits but not for polar qubits. So what we have found
is that for an arbitrary qubit the Hadamard transformations defined by
(9) or (10) do not exist. But for an polar qubit the correct Hadamard 
transform is (9) 
and for an equatorial qubit the correct Hadamard transform is (10).
 
Below we illustrate how definition (9) is at work for polar qubits. 
First notice that we would like to have a unitary transformation that will 
satisfy (9). If we send an unknown `real' qubit through ordinary 
Hadamard gate, we will have 
\begin{eqnarray}
&& \ver \Psi(\theta) \ra \rightarrow {1 \over \sqrt 2} [(\cos \frac{\theta}{2}
+ \sin \frac{\theta}{2}) \ver 0 \ra +( \cos \frac{\theta}{2} - 
\sin \frac{\theta}{2}) \ver 1 \ra] \nonumber\\
&& \ver {\bar \Psi}(\theta) \ra \rightarrow {1 \over \sqrt 2} 
[(\cos \frac{\theta}{2} -\sin \frac{\theta}{2}) \ver 0 \ra - 
( \cos \frac{\theta}{2} + \sin \frac{\theta}{2})\ver 1 \ra ].
\end{eqnarray}
Ideally, we should have obtained 
\begin{eqnarray}
&& \ver \Psi(\theta) \ra \rightarrow 
{1 \over \sqrt 2}[
(\cos \frac{\theta}{2} - \sin \frac{\theta}{2}) 
\ver 0 \ra + (\cos \frac{\theta}{2} + \sin \frac{\theta}{2})\ver 1 \ra] 
\nonumber\\
&& \ver {\bar \Psi}(\theta) \ra \rightarrow 
{1 \over \sqrt 2}[(\cos \frac{\theta}{2} + \sin \frac{\theta}{2}) 
\ver 0 \ra + (\sin \frac{\theta}{2} - \cos \frac{\theta}{2})\ver 1 \ra ]. 
\end{eqnarray}
The actual and the ideal states are different. Hence the ordinary Hadamard gate
cannot be used to create (18). But the desired unitary 
transformation is not difficult to find and is given by the original 
Hadamard matrix times the Pauli spin matrix $\sigma_x$,
i.e., the Hadamard transformation for polar qubits 
is given by $H_P= \sigma_x H = \frac{1}{\sqrt 2} \left( 
\begin{array}{rr} 1 & -1 \\  1 & 1 \end{array} \right)$.
This will create an equal superposition of any arbitrary `real' qubit and 
its complement, i.e., the action of $H_P$ on $ \ver \Psi(\theta) \ra$ will give
$ {1 \over \sqrt 2} (\ver \Psi(\theta) \ra +  \ver {\bar \Psi}(\theta)\ra )$ 
and on $\ver {\bar \Psi}(\theta) \ra$ will give
${1 \over \sqrt 2}(\ver \Psi(\theta) \ra -  \ver {\bar \Psi}(\theta)\ra )$, 
up to an overall minus sign in the later case.

Similarly, one can find a unitary Hadamard gate for an equatorial qubit that
satisfies (10).  If we send $|\Psi(\phi) \ra = H(\cos \frac{\phi}{2} 
\ver 0 \ra - i \sin \frac{\phi}{2} \ver 1 \ra)$ through (10) we have 
\begin{equation}
|\Psi(\phi) \ra \rightarrow \frac{1}{2} [(1+i) e^{i\phi/2} \ver 0\ra + (1-i) e^{-i\phi/2} \ver 1 \ra ]
\end{equation}
and if we send $|{\bar \Psi}(\phi) \ra = H(i \sin \frac{\phi}{2} \ver 0 \ra
- \cos \frac{\phi}{2} \ver 1 \ra )$ through (10) we have 
\begin{equation}
|{\bar \Psi}(\phi) \ra \rightarrow \frac{1}{2} [(1+i) e^{i\phi/2} \ver 0\ra - (1-i) e^{-i\phi/2} \ver 1 \ra ]
\end{equation}
The desired Hadamard gate that will do the above job is given by
$H_E = \frac{1}{\sqrt 2} \left( 
\begin{array}{rr} 1+i & 0 \\  0 & 1-i \end{array} \right)$.
This will create an equal superposition of any arbitrary equatorial qubit and 
its complement, i.e., the action of $H_E$ on $ \ver \Psi(\phi) \ra$ will give
$ {1 \over \sqrt 2} (\ver \Psi(\phi) \ra + i \ver {\bar \Psi}(\phi)\ra )$ 
and on $\ver {\bar \Psi}(\phi) \ra$ will give
${1 \over \sqrt 2}(i \ver \Psi(\phi) \ra + \ver {\bar \Psi}(\phi)\ra )$.

One can also ask if it is possible to create unequal superposition of an
unknown qubit with its complement state? If such a device exist then 
we would have
\begin{eqnarray}
&& \ver \Psi \ra \rightarrow  a \ver \Psi \ra +  b \ver 
{\bar \Psi} \ra) \nonumber\\ 
&& \ver {\bar \Psi}  \ra \rightarrow b^* \ver \Psi  \ra - a^* \ver 
{\bar \Psi}  \ra ), 
\end{eqnarray}
where $a, b$ are {\em known} complex numbers and $|a|^2 + |b|^2=1$.
Using unitarity one can show that the above gate cannot exists.
However, if a qubit is chosen from the polar circle on the Bloch sphere
and if $a,b$ are real, then it is possible to create unequal superposition
of a state with its complement. We know that if a qubit is in
$\ver 0 \ra$ or $\ver 1 \ra$ then one create 
$\ver 0 \ra \rightarrow  a \ver 0 \ra +  b \ver 1 \ra$ and 
$ \ver 1 \ra \rightarrow b \ver 0 \ra - a \ver 1  \ra$ by applying a known
unitary transformation $U=  
\left( \begin{array}{rr} a & b \\  b & -a \end{array} \right)$.
One can check that if we apply 
$U_G= \left( 
\begin{array}{rr} a & -b \\  b & a \end{array} \right)$
to $\ver \Psi(\theta) \ra$, it will give 
$  a \ver \Psi(\theta) \ra +  b \ver {\bar \Psi}(\theta) \ra$ and
to $\ver {\bar \Psi}(\theta) \ra$ will give $ b \ver \Psi(\theta)  \ra 
- a \ver {\bar \Psi }(\theta) \ra$ up to an over all minus sign in 
the later case.
The amplitudes $a,b$ in unequal superposition has to be real, otherwise
the gate will not be `universal' for real qubits. That is, when applied to
two distinct arbitrary qubits, it will not preserve the inner product.
To see this, let $\{\ver \Psi(\theta_1) \ra, \ver \Psi(\theta_2) \ra \}$ 
be two non-orthogonal states and
$\{\ver {\bar \Psi}(\theta_1) \ra, \ver {\bar \Psi}(\theta_2) \ra \}$ be 
their complement states. If the gate has to be universal, it should work 
for all inputs. Suppose $a,b$ are complex, then $\ver \Psi_1(\theta) 
\ra \rightarrow  a \ver \Psi_1(\theta) \ra +  b \ver {\bar \Psi_1}(\theta) 
\ra)$ and $\ver  \Psi_2(\theta)  \ra \rightarrow a \ver \Psi_2(\theta)  \ra + 
b \ver \Psi_2(\theta)  \ra )$. Taking the inner product, we have 
\begin{eqnarray}
\la \Psi_1(\theta) \ver \Psi_2(\theta) \ra \rightarrow  
\la \Psi_1(\theta) \ver \Psi_2(\theta) \ra
 + (a^* b - a b^*) \la \Psi_1(\theta) \ver {\bar \Psi_2}(\theta) \ra,
\end{eqnarray}
where we have used the condition (15).
Similarly, by taking the inner product of 
$\la {\bar \Psi_1}(\theta) \ver {\bar \Psi_2}(\theta) \ra$ we can check 
that it will not preserve the inner product  unless $a,b$ are real. 
This shows that for unequal superpositions of polar qubit with its 
complement state to hold the amplitudes in the superposition should be real.
In an analogous manner one can find transformations for equatorial qubits
also. 

Thus, single qubit gates such as Hadamard and unitary gates 
cannot be designed in an universal manner. The surprising thing is that 
linearity does not allow linear superposition of an unknown qubit with 
its complement!

\section{Non existence of CNOT gate for unknown qubits}

Next, we briefly come to another important gate, namely, the CNOT gate
which is one of the gates needed for universal quantum computation. 
In this section we discuss why it is impossible to design a CNOT gate 
for two qubits that have been prepared in some unknown state.
This is a two-qubit gate 
and takes $\ver 0 \ra \ver 0 \ra \rightarrow \ver 0 \ra \ver 0 \ra, 
\ver 0 \ra \ver 1 \ra \rightarrow \ver 0 \ra \ver 1 \ra,
\ver 1 \ra \ver 0 \ra \rightarrow \ver 1 \ra \ver 1 \ra$ and 
$\ver 1 \ra \ver 1 \ra \rightarrow \ver 1 \ra \ver 0 \ra$.
It flips the second bit if and only if the first qubit is in the state 
$|1\ra$, otherwise  it does nothing. 

One can ask: Does there exist
a CNOT gate for arbitrary two-qubits that will take
\begin{eqnarray}
&& \ver \Psi \ra \ver \Psi \ra \rightarrow \ver \Psi   \ra 
\ver \Psi   \ra,~~~\ver \Psi   \ra \ver {\bar \Psi}
 \ra \rightarrow \ver \Psi   \ra \ver {\bar \Psi}   \ra, \nonumber\\
&& \ver {\bar \Psi}   \ra \ver \Psi   \ra \rightarrow 
\ver {\bar \Psi}   \ra \ver {\bar \Psi}   \ra, ~~~
\ver {\bar \Psi}   \ra \ver {\bar \Psi}   \ra \rightarrow 
\ver {\bar \Psi}   \ra \ver \Psi \ra.
\end{eqnarray}
Again using linearity it can be easily shown that this gate does not exists. 
Physically, this impossibility can be traced to the fact that
CNOT gate measures the first qubit and flips the second one iff the 
first qubit is in the state $\ver {\bar \Psi} \ra$. As we know, 
measuring an unknown qubit without disturbing it, is impossible 
\cite{cf}. Hence one cannot design an universal CNOT gate for all 
qubits. Alternately, the CNOT operator for two qubits in orthogonal 
states given by
\begin{eqnarray}
U_{\rm {CNOT}}^{(0,1)} =  |0 \ra \la 0| \otimes I + 
|1\ra \la 1| \otimes \sigma_x
\end{eqnarray}
cannot be used for arbitrary qubits. Because the desired CNOT operator 
for two unknown qubits would be
given by
\begin{eqnarray}
U_{\rm {CNOT}}^{\Psi,{\bar \Psi} }
 = |\Psi\ra\la \Psi| \otimes I +
|{\bar \Psi}\ra\la {\bar \Psi}| \otimes  \sigma_x(\alpha, \beta) ,
\end{eqnarray}
where $\sigma_x(\alpha, \beta) = (\ver \Psi \ra \la {\bar \Psi} \ver
+ \ver {\bar \Psi} \ra \la  \Psi \ver )$ and this cannot be designed 
without prior knowledge of the amplitudes.
(In fact, other two Pauli matrices in unknown basis such as
$\sigma_y(\alpha, \beta) = -i(\ver \Psi \ra \la {\bar \Psi} \ver
- \ver {\bar \Psi} \ra \la \Psi \ver ),
\sigma_z(\alpha, \beta) = (\ver \Psi \ra \la \Psi \ver
- \ver {\bar \Psi} \ra \la {\bar \Psi } \ver )$ are also impossible 
to measure.)
Thus unknowability of a single quantum rules out the existence of CNOT gate.
Similarly, one can also rule out CCNOT (double-CNOT) and nCNOT (multi-CNOT)
gates for unknown qubits.

\section{Conclusions}

Before concluding we briefly mention the 
implications of the well known limitations and the ones 
discovered in this paper on the future design of quantum computers. 

We suggest that the general limitations, impossibility of designing 
Hadamard gate, unitary logic gate and CNOT gate for 
arbitrary qubits can have some serious implications. 
In a classical computer physical laws do not impose any limitation 
to perform various logical operations
such as NOT, AND, XOR, FANOUT (cloning) and FAN-IN (deleting). Moreover,
arbitrary classical operations can be generated through one bit gate 
such as a NOT and a two-bit gate such as a XOR. In quantum world 
information is stored in superposed states and that makes it completely 
different from classical information. For example, perfect cloning and 
deleting are not allowed 
operations in a quantum computer. Nevertheless, it is well 
known that one bit and two-bit unitary gates are universal for quantum 
computation. However, the limitations on the
one bit and two bit gates suggest that, perhaps, we need to revise 
our understanding about universality of quantum computation. In the 
light of present work
it may be said that even though one qubit gate (an example being a 
Hadamard) and two qubit gate such as a CNOT are universal with respect to 
designing arbitrary unitary operators, they themselves are not universal 
with respect to states. In classical computer these gates are universal
with respect to operations as well as physical states on which information
is stored. But in a quantum computer it is not so. 
In future one would like to investigate further the implications of these
fundamental limitations in quantum information.

In conclusion, we have argued that the impossibility of 
producing a copy and a complement copy are special 
cases of the general limitation. We proved that {\em universal} Hadamard 
and unitary logic operations cannot be performed exactly on arbitrary 
unknown qubits for creating equal and unequal superpositions. The linear 
superposition which 
is at the heart of quantum mechanics, that itself cannot be created for 
a single quantum in an unknown basis. However, if a qubit 
is chosen from polar or equatorial great circle on a Bloch sphere then 
one can design these logic operations by suitably defining the 
transformations. We also discussed why we cannot design a CNOT gate for unknown
qubits. Future avenue of exploration lies in designing universal, approximate 
and optimal general transformations, Hadamard and CNOT gates for 
arbitrary qubits in the spirit of universal estimation \cite{mp}, 
cloning \cite{buzek,gm,rfw} and universal manipulation of qubits 
\cite{bhw,hs,hs1}. Also one can try to realize these impossible 
operations in a probabilistic but in an exact manner analogous to the 
probabilistic cloning \cite{dg}, novel cloning \cite{arun} and 
probabilistic deleting operations \cite{fzy,qiu,fgwz}. In addition, 
one may try to extend these limitations 
and possible operations for higher dimensional and continuous 
variable quantum systems.

\vskip .5cm
{\bf Acknowledgements:} I wish to thank S. Bose, S. L. Braunstein, N. Gisin, 
and various others for very useful discussions on these ideas at several 
stages in last two years. Nevertheless, I am responsible if there are any 
unavoidable errors in the presentation of these ideas.



\begin{thebibliography}{99}

\bibitem{rj} R. Jozsa, {\it Geometric Issues in Foundations of Science},
edited by S. Huggett (Oxford University Press, Oxford, 1997).

\bibitem{wz} W. K. Wootters and W. H. Zurek, Nature {\bf 299}, 802 (1982).

\bibitem{dd} D. Dieks, Phys. Lett. A {\bf 92}, 271 (1982)

\bibitem{hpy} H. P. Yuen, Phys. Lett. A {\bf 113}, 405 (1986).

\bibitem{pb} A. K. Pati and S. L. Braunstein, Nature {\bf 404},  164 (2000)

\bibitem{whz} W. H. Zurek, Nature, {\bf 404}, 40 (2000).

\bibitem{akp} A. K. Pati, (unpublished notes 1998) 

\bibitem{bhw} V. Buzek, M. Hillery and R. F. Werner, Phys. Rev. A 
{\bf 60}, R2626 (1999).

\bibitem{gp} N. Gisin and S. Popescu, Phys. Rev. Lett. 
{\bf 83}, 432 (1999).

\bibitem{jozsa} R. Jozsa, LANL Report, quant-ph/0204153 (2002).

\bibitem{cb} C. Bennett, G. Brassard, C. Crepeau, R. Jozsa, A. Peres,
 and W. K. Wooters, Phys. Rev. Lett. {\bf 70}, 1895 (1993).

\bibitem{akp1} A. K. Pati, Pramana- J. of Phys. {\bf 59}, 217 (2002).

\bibitem{land} R. Landauer, IBM J. Res. Develop. {\bf 5}, 183 (1961).

\bibitem{chb} C. H. Bennett, Int. J. Theor. Phys. {\bf 21}, 905 (1982). 

\bibitem{nc} M. A. Nielsen and I. L. Chuang, Quantum Computation and
Quantum Information, {\it Cambridge University Press}, Cambridge, 2000.

\bibitem{akp2} A. K. Pati, Phys. Rev. A  {\bf 63}, 014302 (2001)

\bibitem{sb} S. Ghosh, A. Roy, and U. Sen, Phys. Rev. A 
{\bf 63}, 014301-1 (2001).

\bibitem{cf} C. A. Fuchs, LANL Report, quant-ph/9611006 (1996).

\bibitem{mp} S. Massar and S. Popescu, Phys. Rev. Lett. {\bf 74}, 1259 (1995).

\bibitem{buzek} V. Buzek and M. Hillery, Phys. Rev. A {\bf 54}, 1844 (1996).

\bibitem{gm} N. Gisin and S. Massar, Phys. Rev. Lett. {\bf 79}, 2153 (1997).

\bibitem{rfw} R. F. Werner, Phys. Rev. A {\bf 60}, 1827 (1998).

\bibitem{hs} L. Hardy and D. D. Song, Phys. Rev. A {\bf 63}, 032301 (2001).

\bibitem{hs1} L. Hardy and D. D. Song, Phys. Rev. A {\bf 63}, 032304 (2001).

\bibitem{dg} L. M. Duan and G. C. Guo, Phys. Rev. Lett. {\bf 80}, 4999 (1998).

\bibitem{arun} A. K. Pati, Phys. Rev. Lett. {\bf 83}, 
2849 (1999).

\bibitem{fzy} Y. Feng, S. Zhang and M. Yim, Phys. Rev. A {\bf 65}, 
042324 (2002).

\bibitem{qiu} D. Qiu, Phys. Rev. A {\bf 65}, 052303 (2002).

\bibitem{fgwz} J. Feng, Y. F. Gao, J. S. Wang, and M. S. Zhan,
Phys. Rev. A {\bf 65}, 052311 (2002).

\end{thebibliography}
\end{document}